# Mid-infrared 2D nonredundant optical phased array of mirror emitters in an InGaAs/InP platform

Jason Midkiff[1], Po-Yu Hsiao[2], Patrick T. Camp[2], and Ray T. Chen[1,2,*]

[1]*Omega Optics, Inc., Austin, TX 78757, USA*
[2]*Department of Electrical and Computer Engineering, The University of Texas at Austin, Austin, TX 78758, USA*
*chenrt@austin.utexas.edu

## Abstract

The extension of photonic technologies such as lidar and free-space optical communications from the traditional visible and near-infrared wavelengths to longer wavelengths can improve performance in adverse environments such as haze, fog, smoke, or strong solar background. Non-mechanical beam steerers will be a critical component of the low size, weight, and power modules needed for the portable or unmanned systems deployed in these environments. In this work, we demonstrate the first 2D optical phased array for non-mechanical beam steering in the mid-infrared spectral region. We combine a total-internal-reflection mirror emitter with a nonredundant array of 30 elements to carry out 2D beam steering at a single wavelength of 4.6 μm. The experiment yielded ~600 resolvable far-field points, with ~2400 points over a 28° × 28° field of view calculated theoretically. Moreover, the device was fabricated in a passive InGaAs/InP platform, contributing another advance in the ongoing development of quantum cascade laser-based photonic integration.

## 1. Introduction

Optical beam forming and steering are critical aspects of applications such as lidar and free-space optical communications [1]. Conventional methods of beam forming and steering are typically mechanical or liquid crystal based. However, motivated by gains in size, weight, and power, electrically-tuned chip-scale solutions have advanced rapidly in recent years [2]. In particular, optical phased array (OPA) technology has seen momentous performance enhancements. Various architectures and physics have been employed to achieve higher-resolution wider-angle steering with improved optical and electrical efficiencies [3]. The majority of this work has been carried out in the near-infrared spectral region, especially in the telecom band, due to the wide availability of affordable sources and detectors and the convenience of mature silicon-based platforms. Conversely, in the mid-infrared spectral region, the limited availability of these elements has slowed progress. Mid-infrared (mid-IR) wavelengths, however, can provide the advantages of lower scattering from atmospheric aerosols and lower background solar noise in comparison to the near-infrared, and thus improve signal-to-noise ratios in certain adverse environments [4]. Hence, translation of the advances in chip-scale OPA technology from near-IR to mid-IR platforms is needed.

The OPA is based on a series of coherent emitting elements whose phases are manipulated to control the superposition of their waves in the far field, thus forming and steering beams of light. The primary lobe of constructive interference is the beam of interest, while other higher order lobes called grating lobes may also exist but are undesirable, since they limit the unambiguous steering range, also known as the field of view (FOV). The FOV, together with the beam width BW (taken at the full-width-half-maximum), determine the important figure of merit for resolution the number of resolvable points: #RP = FOV/BW. The FOV is improved by reducing the inter-element spacing, which may be limited by the size of the emitting elements or by the evanescent coupling between waveguides. The beam width, on the other hand, is benefited by increasing the full aperture size, i.e., by increasing the number of elements. Thus, conventionally, a large number of resolvable points across a wide FOV implies tight packing of numerous emitting elements. Though periodic 1D arrays may manage this packing density with some tradeoffs [5, 6], various works have demonstrated the more favorable approach of using sparse aperiodic arrays to overcome the complexity. Through optimization of the elements' positions, grating lobes can be suppressed or the beam sized reduced, at a minimal cost in the peak side-lobe level (PSLL). The approach has been used successfully with 1D OPAs to achieve 1000s of resolvable points [7, 8].

For 2D OPAs, though, sparsity is especially pertinent since the minimal grid spacing may also be limited by the waveguide routing footprint. Indeed, 2D arrays generally use (uniform) grid spacings of several wavelengths, whether or not fully-populated (i.e., whether or not there exists an emitter at each and every grid point), which is based on the performance strategy. For example, fully-populated grids, which provide the best PSLLs, have necessitated spacings of ~6–7$\lambda$ to accommodate the waveguide routings at the expense of FOV [9–11]. On the other hand, a partially-populated grid achieved a spacing of 3.6$\lambda$ through the use of a genetic algorithm to balance the waveguide routing footprint, resolution, and PSLL [12]. Finally, a spacing of 9.7$\lambda$ was used in a minimally-populated grid, through the concept of a nonredundant array, with the aim of maximizing the aperture size and in turn the resolution [13].

Translating these advances in OPA technology to the mid-IR spectral region is not straightforward, due to the diminishing transparency of silicon dioxide beyond $\lambda \approx 4$ μm. Large core waveguides [14] or material platforms such as silicon-on-sapphire [15] or germanium-on-silicon [16] can be used to extend the transparency window to longer wavelengths. However, hybrid or heterogeneous laser coupling to these platforms will typically be low efficiency or thermally unstable (due to the large amount of heat that integrated QCLs will spread across the bonded interface). The more attractive option is the use of an InP-based platform capable of monolithic integration with QCLs, such as the lattice-matched $In_{0.53}Ga_{0.47}As$/InP platform. In this case, active and passive waveguides are fabricated on the same substrate, without bonding interfaces, permitting high-efficiency coupling and thermal robustness. Indeed, various works have demonstrated the suitability of the lattice-matched InGaAs/InP passive waveguide platform, including low passive losses and efficient coupling schemes with QCLs [17–20]. In our previous work, we demonstrated the first 1D OPA in this platform [21].

In this work, adopting the nonredundant array concept and continuing to employ the passive InGaAs/InP waveguide platform, we demonstrate for the first time a 2D OPA operating at a single mid-IR wavelength of 4.6 μm. The basic architecture of the 2D OPA chip is shown in Figure 1 and consists of three main sections: a power distribution tree, phase shifters, and the emitter array. Each of the $N$ channels possesses an independent phase shifter, and $N$ emitters are distributed in a one-element-per-each-row-and-column fashion over an $N \times N$ uniform grid. The output beam is steered by properly aligning the individual emitter wavefronts via phase shifting.

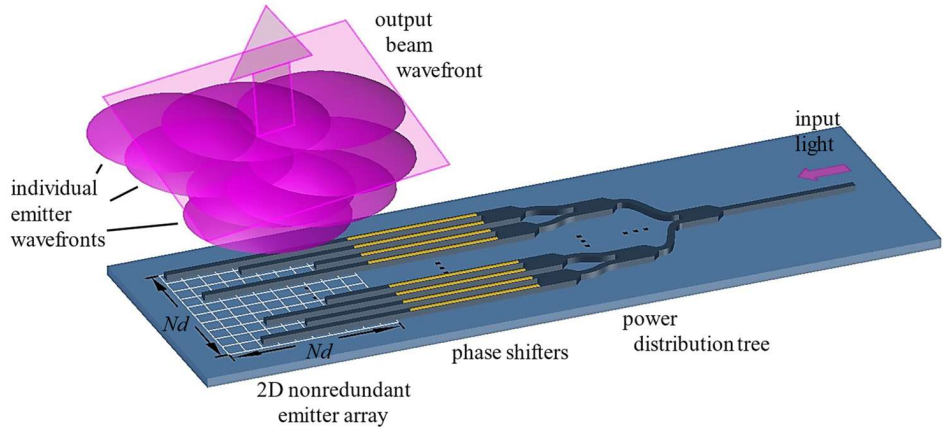

Figure 1: The basic architecture of our 2D nonredundant OPA. The input light is distributed across $N$ channels, each with an independent phase shifter, and $N$ emitters are distributed one-element-per-each-row-and-column fashion over an $N \times N$ uniform grid. Proper phase shifting aligns the individual emitter wavefronts to steer the output beam.

## 2. Emitting elements

To reduce the grid spacing and improve the FOV (by increasing the angle between the grating lobes and the main lobe as described in the following section), we opt for a one-element-per-column approach and implement a waveguide total-internal-reflection (TIR) mirror emitter [21]. The more common grating-based emitter typically occupies a lateral footprint of at least a couple wavelengths [9, 10]. The TIR mirror emitter, on the other hand, requires no extension of the waveguide material. Rather, the footprint is based solely on the details of the etch process. In this case, with the waveguide width less than $1\lambda$, we can reduce the grid spacing to $2\lambda$. Figure 2 shows details of our waveguide mirror emitter and its simulation characteristics. The single-mode ridge waveguide design uses an 850-nm-thick InGaAs core within an InP cladding. (There also exists a 150-nm-thick InGaAs layer within the top cladding as an etch stop in other processes.) The waveguide width $w$ and depth into the substrate $d$ are adjustable through the fabrication process. At the end of the waveguide, the TIR condition is fulfilled from a reverse beveled facet of angle less than ~65°. Upwards transmission is maximum for a bevel angle of 45°. Dry etching methods exist to produce such an angle [23]. However, in general, dry etching alone leaves some degree of surface roughness and should be supplemented with a chemical smoothing process (for example, see [19]) to mitigate the adverse effect on reflection. For simplicity in the fabrication process, we employ a crystallographic wet etch that produces near atomic-level smoothness with a single etch. The etchant is a 4% by volume solution of bromine in methanol. For (001) InP, a reverse bevel at an angle of ~55° is formed when the waveguide is oriented along a $\langle \bar{1}10 \rangle$ direction (which is crystallographically distinct from the $\langle 110 \rangle$ directions) [24].

Emission simulations were carried out with commercial software (Ansys Lumerical FDTD) for waveguides of dimensions $w = 4.3$ μm and $d = 1.0$ μm. The results show that for a bevel angle of 55°, upwards transmission from a bare facet is ~20% and ~30% for TE and TM, respectively (Figure 2(e)). However, these transmissions increase by ~10% each for insulator-coated facets. In fact, our fabrication process utilizes a spin-on-glass layer which incidentally coats the mirror facet, and which we choose not to remove for this reason. The simulated far-field envelope derived from TE operation with a coated facet is shown in Figure 2(f), displaying a longitudinal peak emission angle $\theta_x = -25°$ and a 3dB envelope $\Delta\theta_x \times \Delta\theta_y = 63° \times 55°$. For the case of TM operation, the longitudinal peak emission angle $\theta_x = -34°$ and the 3dB envelope $\Delta\theta_x \times \Delta\theta_y = 89° \times 50°$.

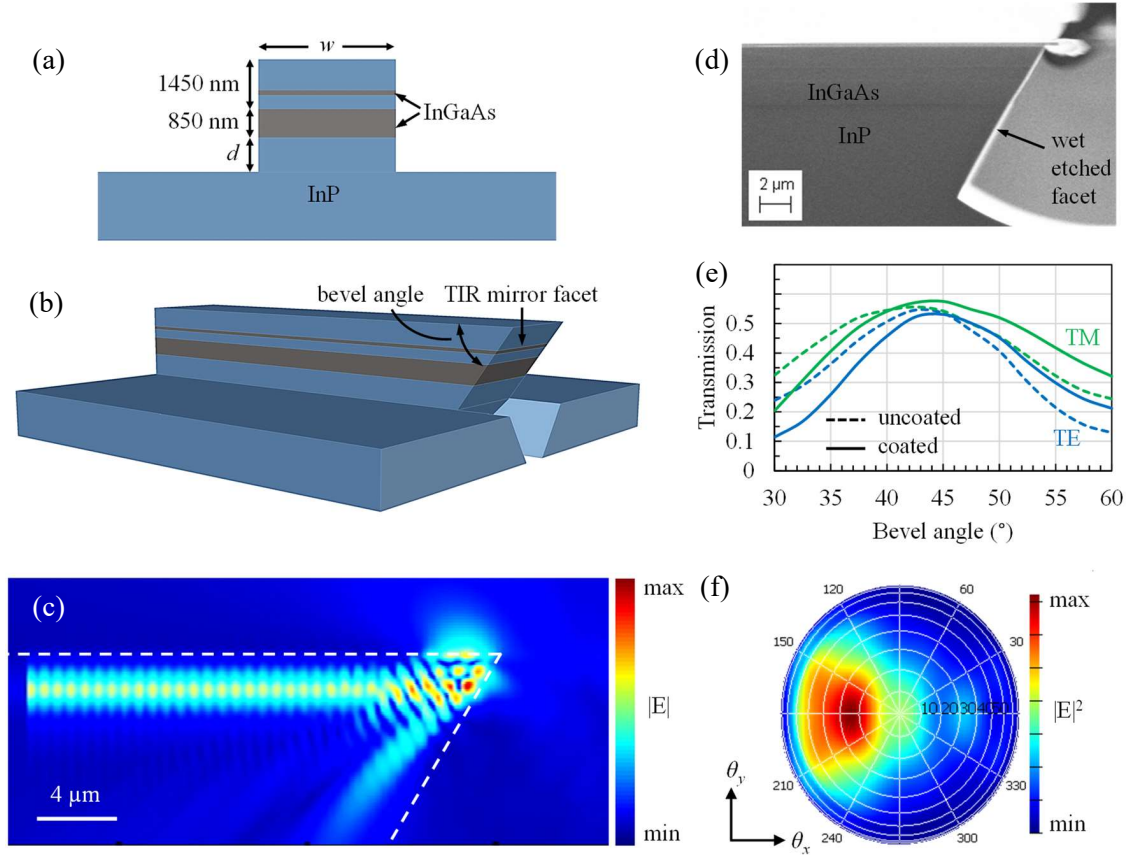

Figure 2: TIR mirror emitter. (a) Waveguide cross-section, (b) perspective illustration of waveguide with TIR mirror facet, (c) simulated side-view of electric field magnitude (TE mode), (d) scanning electron microscope (SEM) image side-view of a wet etched InGaAs/InP material, (e) simulated upwards transmission vs. bevel angle, and (f) simulated far-field electric field intensity (TE mode). The simulations consider a waveguide of $w = 4.3$ µm, $d = 1.0$ µm, a bevel angle of 55°, and a spin-on-glass coated facet.

## 3. Nonredundant array

Following Fraunhofer diffraction theory (by way of the spatial Fourier transform of a near-field pattern and the convolution theorem) the far-field pattern (FFP) or $F(\theta_x, \theta_y)$ of an OPA is proportional to the product of the FFP of a single emitting element $S(\theta_x, \theta_y)$ and the FFP of an array of isotropic emitters $A(\theta_x, \theta_y)$ [2, 3, 13]:

$$F(\theta_x, \theta_y) \propto S(\theta_x, \theta_y) \cdot A(\theta_x, \theta_y). \tag{1}$$

The angles $\theta_x$ and $\theta_y$ are taken with respect to the normal of the aperture, in the longitudinal and lateral directions in the far field, as illustrated in Figure 3. The far-field intensity distribution is $|F(\theta_x, \theta_y)|^2$. The array factor leads to the interference pattern and is determined by the arrangement of the emitting elements and their excitations:

$$A(\theta_x, \theta_y) = \sum_n A_n \exp(-jk(x_n \sin\theta_x + y_n \sin\theta_y) + j\varphi_n) \tag{2}$$

where $k = 2\pi/\lambda$ is the wavevector, $\lambda$ is the free-space wavelength, and $(x_n, y_n)$ are the coordinates of the $n$th element with excitation $A_n \exp(j\varphi_n)$. Beam steering to an angle $(\theta_{xs}, \theta_{ys})$ is effected by adjusting the phases $\varphi_n$ to give 0 in the exponential of Eq. (2), i.e., by satisfying

$$\varphi_n = k(x_n \sin\theta_{xs} + y_n \sin\theta_{ys}) \tag{3}$$

for all $n$ elements.

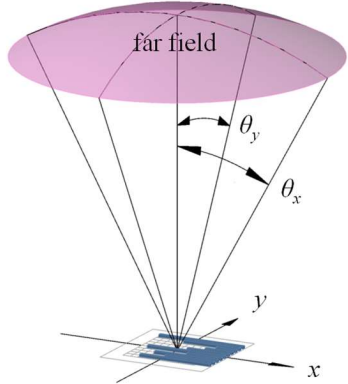

Figure 3: Relation between device spatial coordinates $(x, y)$ and far-field angles $(\theta_x, \theta_y)$.

Eq. (2) is completely general with respect to the elements' positions. An array of elements on a uniform grid, whether fully or partially populated, will exhibit far-field intensity maxima determined by the grid periodicity:

$$\sin\theta_x = \frac{m_x \lambda}{d}, \quad \sin\theta_y = \frac{m_y \lambda}{d}, \tag{4}$$

where $m_x$ and $m_y$ (= ..., -1, 0, 1, ...) are the diffraction orders in the relevant directions and $d$ is the grid spacing, which is equal to the smallest inter-element spacing. In other words, the grating lobes' locations are determined by the smallest inter-element spacing. When the main beam is steered to $(\theta_{xs}, \theta_{ys})$, the grating lobes will move accordingly:

$$\sin\theta_x - \sin\theta_{xs} = \frac{m_x\lambda}{d}, \quad \sin\theta_y - \sin\theta_{ys} = \frac{m_y\lambda}{d}. \tag{5}$$

The FOV is determined by the unambiguous steering range, i.e., the range over which the grating lobes don't appear while steering. In either direction, the main and grating lobes are ambiguous when $\theta = -\theta_s$ for $m = -1$, i.e., when

$$\sin(-\theta_s) - \sin\theta_s = \frac{-\lambda}{d}$$

$$-2\sin\theta_s = \frac{-\lambda}{d}$$

$$\sin\theta_s \equiv \sin\theta_{boundary} = \frac{\lambda}{2d}, \tag{6}$$

The FOV is then twice $\theta_{boundary}$ minus the finite beam width at the boundary:

$$\text{FOV} = 2\theta_{boundary} - \text{BW}. \tag{7}$$

The beam width is inversely related to the full aperture size, discussed further below. The final figure of merit is the peak side-lobe level, defined as the ratio of the intensities of the maximum side lobe and the main lobe:

$$\text{PSLL} = \frac{I_{max\ side\ lobe}}{I_{main\ lobe}}. \tag{8}$$

To illustrate the tradeoffs involved in the arrangement of emitting elements, we compare two simple cases, each with an equivalent number of emitting elements and aperture size, but with different grid spacings limited by their elements' sizes. Figure 4(a) presents the first case: a 10-element array on a 5 × 5 grid of spacing $4\lambda$. Figure 4(c) presents the second case: a 10-element array on a 10 × 10 grid of spacing $2\lambda$. The corresponding simulated far-field patterns along one principal axis are shown in Figure 4(b) and (d). Since the aperture size is the same for each case, the beam widths are nearly equal. On the other hand, the halved grid spacing of the second case leads to a doubling of the FOV in each direction, and consequently a quadrupling of the number of (2D) resolvable points. Specifically, for the first case, #RP = FOV/BW = $(12.8°/1.61°)^2 \approx 63$, while for the second case it's $(27.3°/1.71°)^2 \approx 254$. Additionally, for the second case, the smaller element size produces a broader envelope, meaning that the power of the main beam during steering will not decline as quickly as for the first case. The gains in resolution and main beam power come with a trade-off in the peak side-lobe level, which is increased by 2.7 dB. However, it is important to note that the rather poor PSLL of this simple example is mainly due to the small number of elements and is readily improved with an increase of this number as described further below.

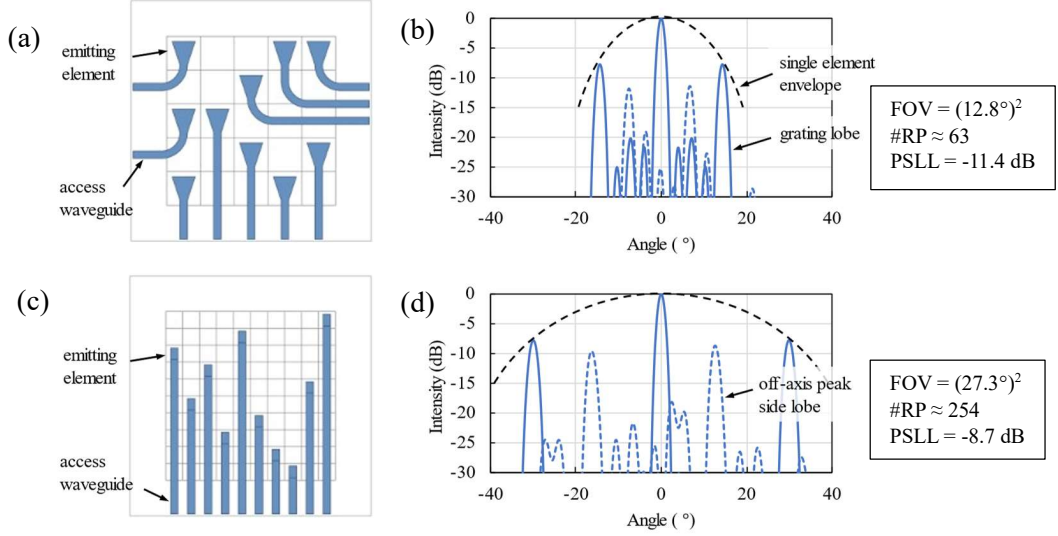

Figure 4: Emitting element distribution conceptual illustration: (a) 10 elements on a 5 × 5 grid of 4$\lambda$ spacing, (b) corresponding simulated far-field pattern along one axis, (c) 10 elements on a 10 × 10 grid of 2$\lambda$ spacing, and (d) corresponding far-field intensity pattern (logarithmic scale) along one axis. Note that the dashed curves show the peak side lobes within the 2D FOVs, which do not lie on the principle axes.

Now we elaborate further on the concept of a nonredundant array. When $N$ elements are arranged on an $N \times N$ grid such that each row and column are populated by exactly one element, mathematically the array is known as a permutation matrix, of which there exist $N!$ distinct permutations [25]. Each permutation will produce a different FFP. Since the FFP is derived from the Fourier transform of the NFP, a narrow beam profile (for a given aperture size) requires that the number of spatial frequencies in the near field be maximized. In other words, the displacement vectors between each pair of elements should be as varied as possible. A permutation matrix in which all displacement vectors are unique is nonredundant, known as a Costas array [26]. Hence, the use of a Costas array in the design of an OPA, in combination with narrow emitters, provides a means of maximizing the resolvable point count (through the narrow beam profile) over a large FOV (through the reduced grid spacing) with simple parallel access waveguides. The example of Figure 4(c) is a Costas OPA.

Furthermore, we note that the trade-off of PSLL for #RP for arrays based on not-fully-populated grids will be less pronounced as the number of elements increases [27–29]. For a Costas OPA, the benefit of increasing the number of elements $N$ on both the BW and PSLL is especially significant. Figure 5 presents the simulated dependence of these characteristics on $N$. The BW decreases because of the increasing aperture size, while the PSLL is reduced through the increased number of destructively interfering beams. Indeed, for large $N$ ($\gtrsim 150$), the PSLL of a Costas array approaches that of a tightly packed uniform array of equivalent $N$.

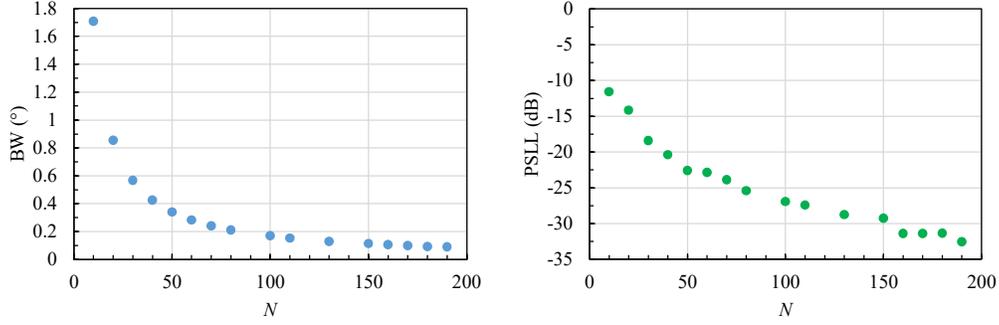

Figure 5: Effect of the number of elements $N$ on the characteristics of a Costas OPA (simulation). (a) Beam width and (b) peak side-lobe level.

Our mid-infrared 2D OPA is a Costas array of 30 elements. The emitting elements' positions were determined through the use of a particle swarm optimization [30] to minimize the beam width subject to the rules of permutation matrices, leading to the uniquely broad spatial frequency spectrum characteristic of the Costas array. The derived array is shown in Figure 6(a), with the corresponding simulated FFP for $(\theta_{xs}, \theta_{ys}) = (0, 0)$ shown in Figure 6(b).

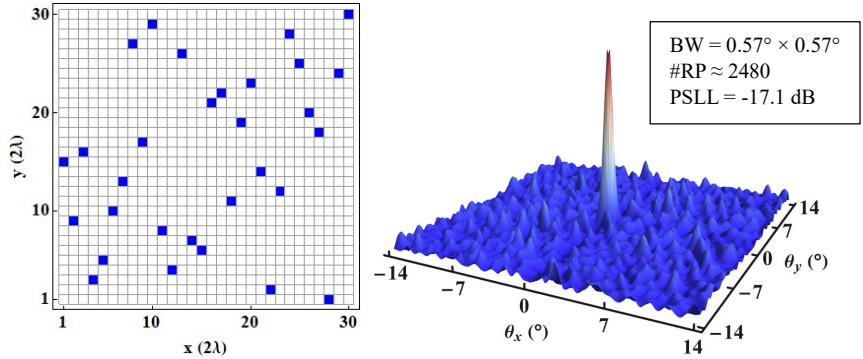

Figure 6: Sparse aperiodic array. (a) Derived 30-element Costas array, and (b) simulated far-field intensity pattern (linear scale) for $(\theta_{xs}, \theta_{ys}) = (0, 0)$ within the FOV.

We have also characterized the beam width and peak side-lobe level as the beam is steered within the field of view. Figure 7 shows the simulation results. Both characteristics degrade to some extent in steering to the edge of the FOV. Each component of the BW increases by just 3% along its corresponding axis. The PSLL increases by up to 2.3 dB along a single axis, and up to 3.1 dB in steering along both axes, i.e., to the corner of the FOV. It should be noted that the increase of the PSLL with steering is due primarily to the decrease in intensity of the main beam, i.e., the denominator of Eq. 8.

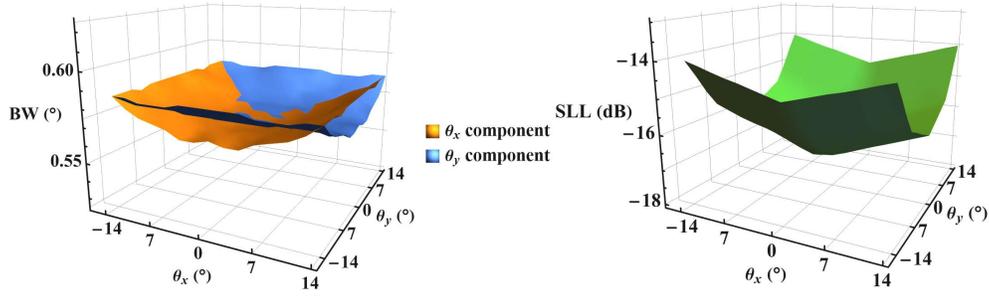

Figure 7. Effects of steering on OPA characteristics. (a) Beam width, and (b) peak side-lobe level.

## 4. Experiment

A schematic illustration of our OPA layout is shown in Figure 8. Light is input through a cleaved waveguide facet (of 20-μm-width due to an inverse taper) and split into 32 separate waveguides of 20-μm-pitch by a 5-level multimode interferometer (MMI) tree. However, due to a testing limitation of 32 probe contacts (= 30 biases and 2 grounds), 30 channels are dedicated to the phased array, while 2 channels are used for laser alignment. As such, 30 thermo-optic phase shifters follow the MMI tree, then the waveguides are fanned in to the 9.2-μm-pitch emitter array. The phase shifting characteristics were evaluated separately as described in the Supplement. Since the TIR mirrors reflect backwards from the normal, the "U-shaped" architecture was implemented to accommodate our experimental setup, which favors emission towards the opposite side of the chip from input coupling. The devices were fabricated in our university facility. Details of the fabrication process can be found in the Supplement. Photos of a fabricated device are shown in Figure 9.

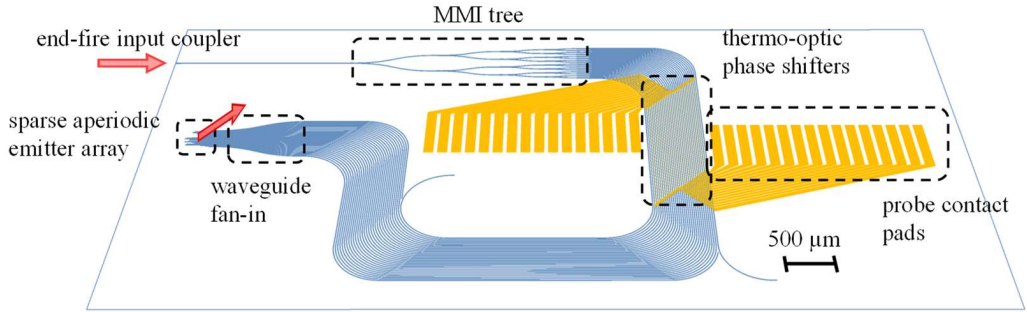

Figure 8. Schematic illustration of OPA layout. The waveguide fan-in reduces the inter-waveguide pitch from 20 μm to $2\lambda = 9.2$ μm.

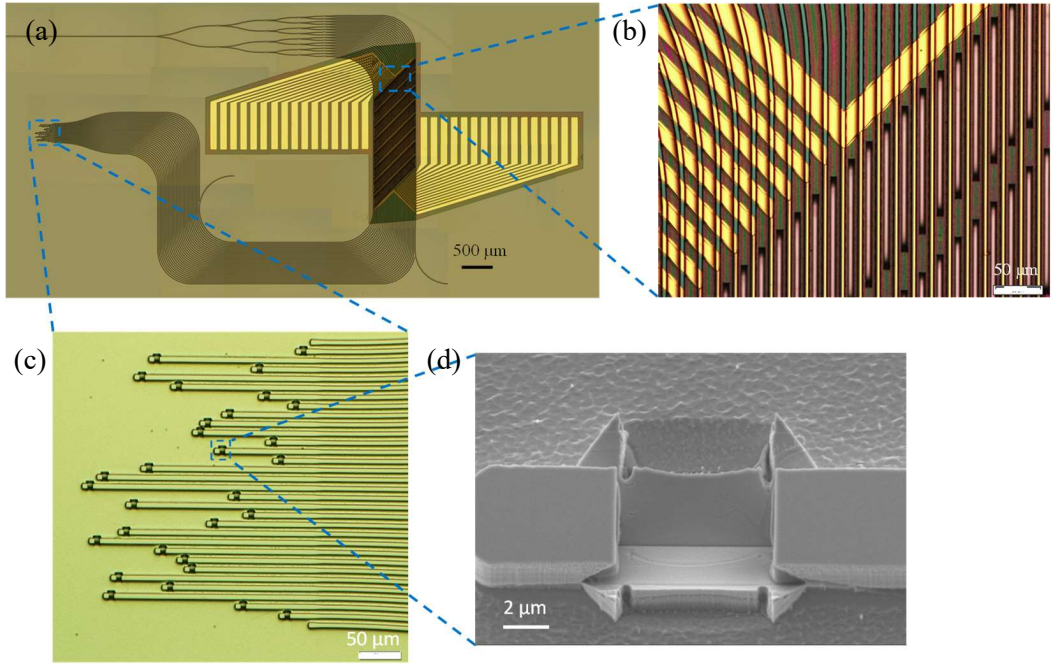

Figure 9. Fabricated OPA device. (a) Stitched photo of full device, (b) magnified region of thermo-optic phase shifter and interconnect metal, (c) magnified emitter array aperture, and (d) SEM image of a single TIR emitter.

Our device testing setup is shown in Figure 10. The OPA chip was mounted to a Peltier cooling block and maintained to ±0.5 °C of room temperature. A pair of 16-contact probes were used to bias the device's phase shifters. A 16-mm-focal-length (at $\lambda$ = 4.6 µm) bi-convex CaF$_2$ lens was used to focus the mid-infrared laser light onto the device's input waveguide facet (in TE polarization), and a Fourier imaging assembly was used to observe the output beam. A schematic illustration of the imager is shown in Figure 10(b). Two 1-inch-diameter 42.9-mm-focal-length (at $\lambda$ = 4.6 µm) plano-convex CaF$_2$ lenses (L1 and L2) formed a 4f correlator that produced a 1× near-field image where a 700-µm-diameter pinhole was positioned for spatial filtering. The pinhole was centered on the emitter array, allowing light from the emitters to pass through while blocking excess uncoupled or substrate-scattered light. A third equivalent lens (L3) was used as the Fourier lens to create the far-field image at a distance of one focal length. The camera was focused on the Fourier plane, where each point corresponds to a far-field steering angle. Finally, we note that the imager assembly was mounted at 33° to the vertical to fit into our optical bench, and that the 1-inch-diameter lenses limited the FOV to a diameter of 23°, i.e., 5° smaller than the device's theoretical FOV.

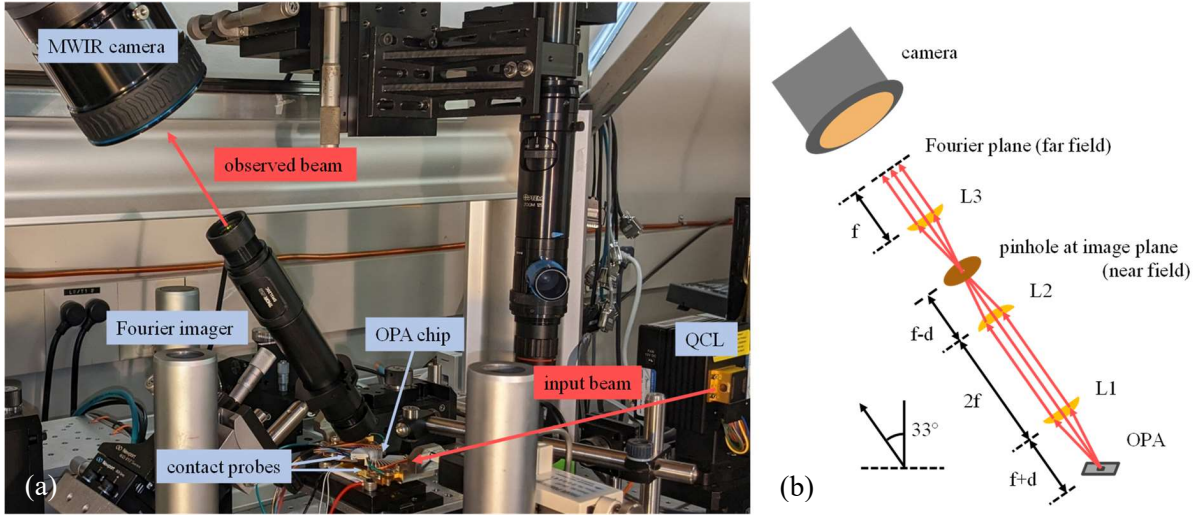

Figure 10. OPA device testing. (a) Photo of the test setup and (b) schematic illustration of the Fourier imaging assembly.

Due to the non-equivalent optical path lengths of the access waveguides, the emitters' phases are initially disordered. Thus, a phase calibration process is required to bring all the emitters into phase coherence. In general, with fully characterized phase shifters, following a single calibration routine, phase gradients can be imposed to steer the beam to an arbitrary far-field point ($\theta_x$, $\theta_y$) [31, 32]. However, to demonstrate steering to just a few arbitrary points we found it easier to carry out the phase calibration process separately for each point, storing the resulting bias conditions for each point in a lookup table. For this purpose, an iterative optimization routine based on a basic hill climbing algorithm was created in NI LabVIEW. The algorithm runs through each channel, adjusting their phases from 0 to $2\pi$, while monitoring the PSLL within the FOV. The phase of each channel is set to the value which yielded a minimum PSLL before proceeding to the successive channel. We found that with the phases discretized into ~20 values and after ~3 full iterations of the routine the PSLL was stable, i.e., further phase discretization or more iterations yielded little to no gain. Various arbitrary steering points were phase calibrated. The 2D far-field patterns of four arbitrary points are shown in Figure 11, annotated with their extracted BWs and PSLLs. Note that the plots are centered at (33°, 0°) due to the imager setup, i.e., at an 8° offset from the simulated emission center at (25°, 0°).

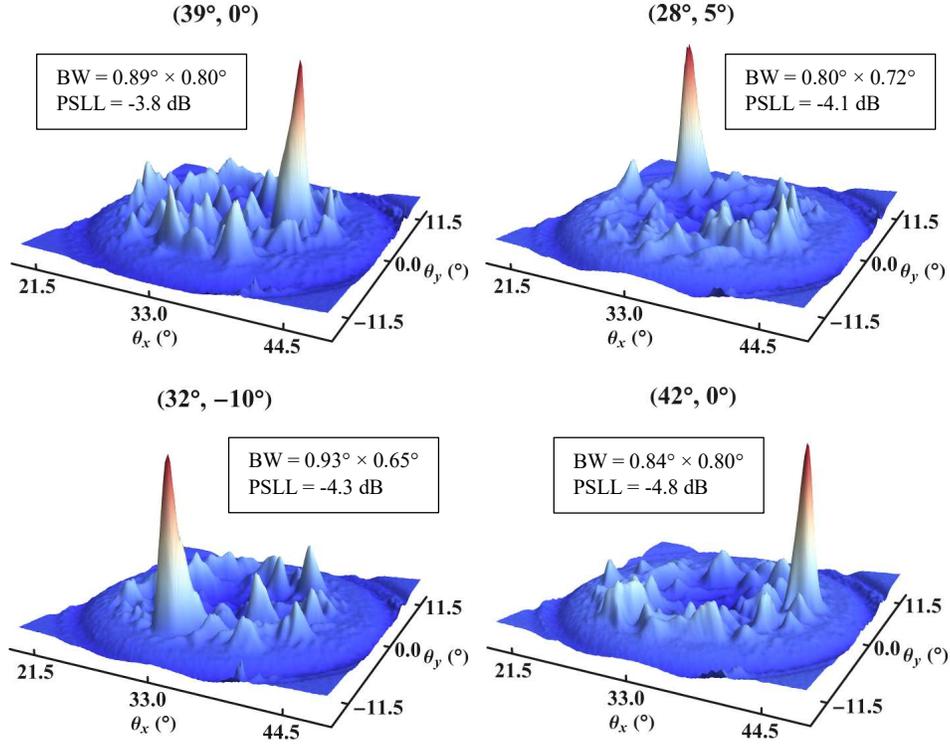

Figure 11. Experimental far-field patterns at four arbitrary steering points.

## 5. Discussion

Table 1 provides the comparison between simulation and experiment. The ideal OPA simulation is the result provided in Figure 6. The non-ideal OPA simulation takes into account the 8° offset in $\theta_x$ as well as estimated channel phase errors in the experiment. These errors are a consequence of the fact that the calibration routine relies on intensity readings for optimization; however, inherent quasi-sinusoidal intensity fluctuations of the QCL gave rise to non-steady readings, impacting the feedback algorithm and the accuracy of phase tuning. For this reason, 20% phase error is set across all channels in the non-ideal simulation. Additionally, four channels were found to be disconnected, so four channels in the simulation were set to have full random phase. Further details of the effects of phase error are presented in the Supplement. It is seen that with the non-idealities in offset and phase error, the beam width and therefore also the number of resolvable points change very little. On the other hand, the peak side-lobe level is raised by nearly 6 dB. The experimental data are presented for both the test-limited FOV (due to the size of the imaging lenses) and the extrapolation to the full device-limited FOV (due to the presence of grating lobes as derived from theory). Despite the non-idealities included in the simulation, the experimental results are still degraded in comparison. The BWs are increased to yield about half the number of resolvable points, and the PSLL is increased by ~7 dB.

Table 1. Comparison between simulation and experimental results of OPA.

|  | Ideal OPA Simulation | Non-ideal OPA Simulation[a] | OPA Experiment, Test-limited FOV[b] | OPA Experiment, Device-limited FOV[c] |
|---|---|---|---|---|
| FOV: | 28.4° × 28.4° | 28.4° × 28.4° | $\pi (11.5°)^2$ | 28.1° × 28.2° |
| BW: | 0.57° × 0.57° | 0.59° × 0.57° | 0.86° × 0.74° | 0.86° × 0.74° |
| #RP: | ≈ 2480 | ≈ 2390 | ≈ 650 | ≈ 1240 |
| PSLL: | -17.1 dB | -11.3 dB | -4.2 dB | -4.2 dB |

[a] Considers the effects of offset steering angle and phase error as described in the text.
[b] Due to the size of the imaging lenses.
[c] Due to the presence of grating lobes as calculated by theory.

One factor not taken into account in the simulation which can have an impact on the experiment is the aberrations in the imaging assembly. The 1-inch-diameter lenses are off-the-shelf components, but the small size is accompanied by spherical aberrations which degrade directly the BW and indirectly the PSLL through added impairment to the phase calibration routine. Customized lenses for a larger-FOV imaging assembly could help to mitigate these experimental challenges and improve the results. However, addressing the more fundamental issue of low signal-to-noise ratio would be a more fruitful endeavor (rendering the imaging assembly unnecessary). The background noise could be reduced by improving both the input and output coupling efficiencies. The input coupling is basically an experimental issue and would be improved by replacing the bi-convex focusing lens with a smaller-focal-length aspherical one. The output coupling efficiency, on the other hand, is a device design issue. The TIR mirrors should be optimized for maximum efficiency, through a more sophisticated fabrication process to create 45° bevels and perhaps the incorporation another component to improve directionality, such as an anti-reflective coating. Finally, an improvement in the phase tuning process to permit more accurate feedback and finer discretization of phases would lead to more accurate calibrations. The main issue here was the oscillating QCL power, which could be stabilized by through a feedback loop to the laser controller.

## 6. Conclusions

Advances in mid-infrared photonics continue to progress at a steady rate. In this work, we have contributed to this advancement through the development of the first mid-infrared 2D optical phased array in an InGaAs/InP waveguiding platform. The nonredundant array concept, and in particular the Costas array, was adopted to maximize the number of resolvable points from $N$ elements on an $N \times N$ grid. For the modest case of $N = 30$, >2000 resolvable points are possible with a peak side-lobe level down to -17 dB. At the same time, we took advantage of the small footprint of TIR mirror emitters to reduce the grid spacing down to $2\lambda$ and achieve a 28° field of view in both dimensions. Testing non-idealities such as low-resolution phase tuning and optical aberrations degraded experimental results somewhat but are readily addressable in next generation devices. Moreover, as $N$ scales to produce more resolvable points, the side-lobe level will also inherently improve. Thus, we expect this work to serve as solid steppingstone in the progress towards a practical mid-IR 2D phased array beam steering technology.

**Funding.** Air Force Research Laboratory (FA9453-22-C-A015).

**Acknowledgement.** Device fabrication was carried out at the University of Texas at Austin's Microelectronic Research Center.

**Disclosures.** The authors declare no conflicts of interest.

See Supplement for supporting content.

# Mid-infrared 2D nonredundant optical phased array of mirror emitters in an InGaAs/InP platform: supplement

Jason Midkiff[1], Po-Yu Hsiao[2], Patrick T. Camp[2], and Ray T. Chen[1,2,*]

[1]*Omega Optics, Inc., Austin, TX 78757, USA*
[2]*Department of Electrical and Computer Engineering, The University of Texas at Austin, Austin, TX 78758, USA*
*chenrt@austin.utexas.edu

**Device Fabrication**

The undoped InP/InGaAs epitaxial layers were grown on an undoped InP substrate by MOCVD at OEpic Semiconductors, Inc. The first step in device fabrication was the formation of the facets that become the TIR mirror emitters. The emitter regions were then defined lithographically in a sacrificial $SiO_2$ hard mask and etched with a bromine-methanol solution. Subsequently, the waveguides were patterned into another sacrificial $SiO_2$ layer and etched by $Cl_2/CH_4/H_2$ ICP-RIE, etching through the core InGaAs layer and into the InP substrate at least 1.0 µm. Next, the waveguides were coated with a spin-on-glass, which served as a spacing between the (later-deposited) metallization and the waveguide and also created a sloped sidewall for continuity in the metallization. The phase shifter heat isolation trenches were then patterned into the spin-on-glass layer and etched by the bromine-methanol solution to a depth of ~10 µm. Next, two metallizations were carried out by deposition and lift-off. The first metallization of Ti/Au (10nm/220nm) served to form the high-resistance 2-µm-wide strip heaters, while the second metallization of Au (400nm) produced the low-resistance interconnects and probe pads. Finally, the spin-on-glass layer was patterned and removed from the non-metallized regions by a sequence of $CHF_3$/Ar RIE and buffered HF wet etch.

**Thermo-optic phase shifters**

The thermo-optic phase shifters consist of 0.22-µm-thick × 2.0-µm-wide gold strip heaters running along the top side of the (spin-on-glass-coated) waveguides. The small cross-section makes them more resistive than the interconnect metal which is 0.60 µm thick and increases in width from 2.0 to 100 µm. The strip heater length is 1660 µm. For the simple architecture of ridge waveguides (with dimensions commensurate of the wavelength 4.6 µm) lined by strip heaters, the 2π-phase-shift power is nearly 500 mW. To reduce this power, heat isolation trenches were implemented. The trenches were created through a crystallographic wet etching process (via a bromine-methanol solution) which undercuts the waveguides. (We point out that the trenches were implemented with lengths of 166 µm and an overall duty cycle of 80% along the length of the phase shifters. I.e., 20% of the phase shifter is left intact for structural integrity.) Mach-Zehnder interferometers (MZIs) were used to measure the phase-shifting performance of the heat-isolated phase shifters. Typically, the trenches reduce the 2π-phase-shift to about 250 mW. However, we also found that if the waveguides were completely undercut (i.e., suspended in air) the 2π-phase-shift power would reduce to less than 100 mW. Figure S1 compares two relevant cases. The first case (Figure S1(a,b)) is that of a device with a typical pair of trenches which come to within ~2 μm of each under the waveguide, but do not coalesce. The second case (Figure S1(c,d)) is that of an exemplary device in which the waveguide is completely undercut.

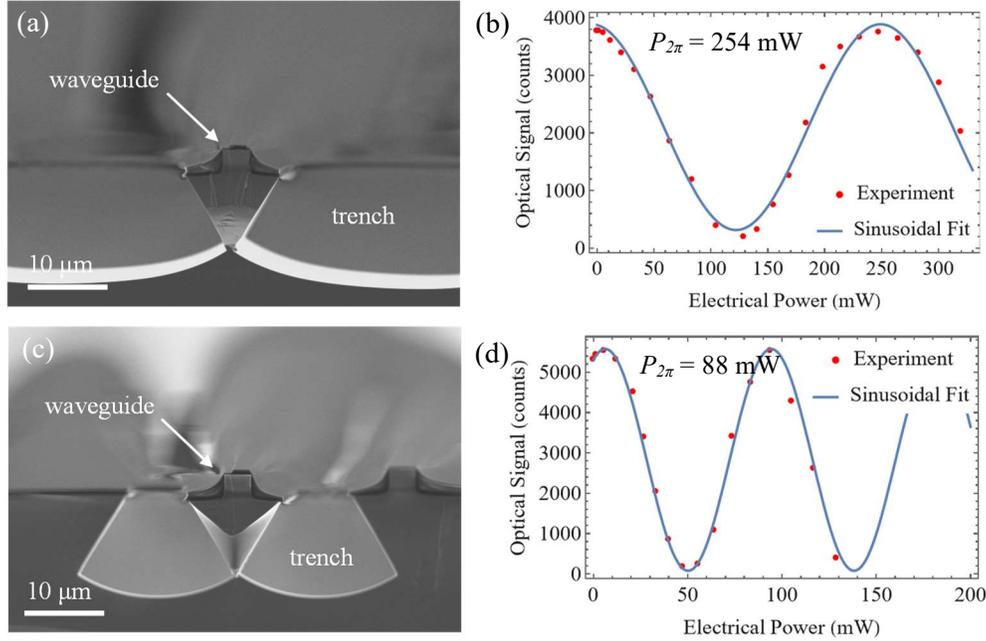

Figure S1: Thermo-optic phase shifter evaluation. (a) SEM cross-section image of a waveguide with a typical heat-isolation trench, (b) corresponding MZI phase-tuning test result, (c) SEM cross-section image of a waveguide completely undercut by the heat-isolation trench, and (d) corresponding MZI phase-tuning test result.

**Effects of Phase Error**

To investigate the effects that residual phase error have on beam formation in the far field, we have simulated the effects of such random error on both the beam widths (BWs) and the peak side-lobe level (PSLL) of our 30-channel sparse aperiodic OPA. Figure S2(a, b) show the effects of systematic phase error, that is to say, that all channels experience some random level of phase error, up to the indicated percentage. For example, 20% phase error refers to the case where each channel may suffer from 0 to 20% deviation from the ideal phase. It is seen that BW is little affected by phase error, staying nearly constant up to ~40% error. PSLL, on the other hand, is more significantly affected, increasing by ~2 dB at 20% error and by ~8 dB at 40% error. Figure S2(c, d) show the effects of disconnected channels, i.e., channels with completely random phase. In this case, both BW and PSLL degrade steadily, with PSLL again suffering more significantly. It is seen that with just two disconnected channels, PSLL can increase by up to ~7 dB.

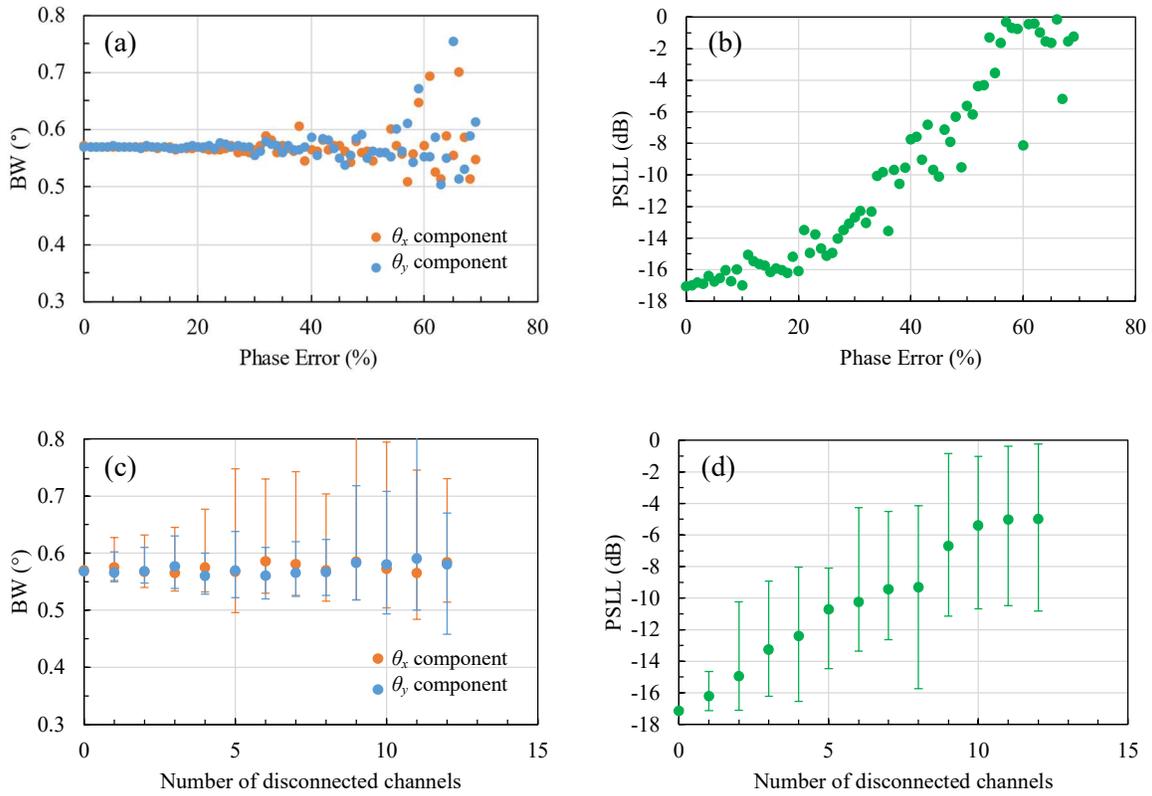

Figure S2: Simulated effects of phase error on BW and PSLL on 30-channel sparse aperiodic OPA. (a) BWs when all channels possess up to indicated percentage of phase error, (b) PSSL when all channels possess up to the indicated percentage of phase error, (c) BWs vs. number of disconnected channels, and (d) PSLL vs. number of disconnected channels (error bars express the range of values acquired across 20 different simulations).

19